\documentclass[twocolumn,aps,prb,superscriptaddress,amsmath,amssymb]{revtex4-2}
\usepackage[breaklinks=true,colorlinks=true,linkcolor=blue,urlcolor=blue,citecolor=blue]{hyperref}
\usepackage{graphicx}
\usepackage{dcolumn}
\usepackage{bm}

\usepackage{siunitx} 
\usepackage{miller} 

\DeclareSIUnit{\bell}{B}
\DeclareSIUnit{\dB}{\deci\bell}

\DeclareSIUnit{\bellwatt}{Bm}
\DeclareSIUnit{\dBm}{\deci\bellwatt}

\begin{document}


\title{Parametric generation of spin waves in nano-scaled magnonic conduits}

\author{Björn Heinz}
\email{bheinz@rhrk.uni-kl.de}
\affiliation{Fachbereich Physik and Landesforschungszentrum OPTIMAS, Technische Universität Kaiserslautern, D-67663 Kaiserslautern, Germany}

\author{Morteza Mohseni}%
\affiliation{Fachbereich Physik and Landesforschungszentrum OPTIMAS, Technische Universität Kaiserslautern, D-67663 Kaiserslautern, Germany}

\author{Akira Lentfert}%
\affiliation{Fachbereich Physik and Landesforschungszentrum OPTIMAS, Technische Universität Kaiserslautern, D-67663 Kaiserslautern, Germany}

\author{Roman Verba}%
\affiliation{Institute of Magnetism, UKR-03142 Kyiv, Ukraine}

\author{Michael Schneider}%
\affiliation{Fachbereich Physik and Landesforschungszentrum OPTIMAS, Technische Universität Kaiserslautern, D-67663 Kaiserslautern, Germany}

\author{Bert Lägel}%
\affiliation{Nano Structuring Center, Technische Universität Kaiserslautern, D-67663 Kaiserslautern, Germany}

\author{Khrystyna Levchenko}%
\affiliation{Faculty of Physics, University of Vienna, A-1090 Wien, Austria}

\author{Thomas Brächer}%
\affiliation{Fachbereich Physik and Landesforschungszentrum OPTIMAS, Technische Universität Kaiserslautern, D-67663 Kaiserslautern, Germany}

\author{Carsten Dubs}%
\affiliation{INNOVENT e.V. Technologieentwicklung, D-07745 Jena, Germany}

\author{Andrii V. Chumak}%
\affiliation{Faculty of Physics, University of Vienna, A-1090 Wien, Austria}

\author{Philipp Pirro}%
\affiliation{Fachbereich Physik and Landesforschungszentrum OPTIMAS, Technische Universität Kaiserslautern, D-67663 Kaiserslautern, Germany}


\begin{abstract}
The research field of magnonics proposes a low-energy wave-logic computation technology based on spin waves to complement the established CMOS technology and provide a basis for emerging unconventional computation architectures.
However, magnetic damping is a limiting factor for all-magnonic logic circuits and multi-device networks, ultimately rendering mechanisms to efficiently manipulate and amplify spin waves a necessity. In this regard, parallel pumping is a versatile tool since it allows to selectively generate and amplify spin waves. While extensively studied in microscopic systems, nano-scaled systems are lacking investigation to assess the feasibility and potential future use of parallel pumping in magnonics. Here, we investigate a longitudinally magnetized \SI{100}{\nano\metre}-wide magnonic nano-conduit using space and time-resolved micro-focused Brillouin-light-scattering spectroscopy. Employing parallel pumping to generate spin waves, we observe that the non-resonant excitation of dipolar spin waves is favored over the resonant excitation of short wavelength exchange spin waves. In addition, we utilize this technique to access the effective spin-wave relaxation time of an individual nano-conduit, observing a large relaxation time up to \SI[separate-uncertainty = true]{115.0(76)}{\nano\second}. Despite the significant decrease of the pumping efficiency in the investigated nano-conduit, a reasonably small threshold is found rendering parallel pumping feasible on the nanoscale.
\end{abstract}

\maketitle


\section{\label{sec:level1}Introduction}
Spin-wave based computing attracts increasing attention in the view of a low-energy computation technology by replacing charge-based binary logic with a wave-based logic utilizing spin waves as the information carrier \cite{Khitun_2010,Mahmoud2_2020,Pirro_2021}. This potentially offers several advantages, e.g., energy efficiency due to the absence of Ohmic losses \cite{kajiwara_2010}, multiple degrees of freedom (frequency and phase) \cite{Schneider_2008}, readily accessible nonlinear mechanisms \cite{Krivosik_2010,Sadovnikov_2016} and a device feature size with a sub-\SI{100}{\nano\metre} scaleability \cite{Heinz_2020}. A variety of spin-wave based devices has already been realized, such as transistors \cite{chumak2014magnon}, majority gates \cite{Talmellieabb4042} or a directional coupler \cite{WangCoupler_2020}, with many more concepts proposed theoretically. For many of these proof-of-concept devices yttrium iron garnet (YIG) is used due to its record low magnetic damping, but the material has been limited to the macro- and microscale for a long time. Recent progress in the fabrication of high quality nanometer thin YIG films \cite{Dubs_2020} and advanced structuring procedures revealed a long-range spin-wave propagation in sub-\SI{100}{\nano\metre}-wide YIG conduits, proving the feasibility of spin-wave computing on the nanoscale \cite{Heinz_2020,Heinz2021}. However, magnetic damping and the fan-out-requirement constitute limiting factors for extended networks and all-magnonic circuits, ultimately rendering spin-wave amplifiers a necessity \cite{Mahmoud2_2020}. Parametric amplification by parallel pumping (PP) has been proven to be a versatile tool in this context, which not only allows for the mere generation, but also for an amplification and manipulation of spin waves \cite{BRACHER20171}. In this process, microwave photons are converted into pairs of spin waves with a well-defined phase relation, providing a frequency- and mode selectivity. In this regard, PP is superior to other approaches of spin-wave amplification, e.g., the spin-transfer-torque amplification of spin waves \cite{ANSTT}, which are often lacking the necessary selectivity. The process of PP has been well studied in microscopic and sub-micrometer systems \cite{PhysRevB.84.094401,BRACHER20171,HwangWire_2021,HwangBlock_2021}, but only theoretically \cite{Haghshenasfard_2016,doi:10.1063/1.5003660,VerbaPump_2018} on the nanoscale or experimentally in fully quantized structures, e.g., nano-discs \cite{Guo_Discs}. However, in open systems, e.g., conduits, the dissipation of propagating spin waves from the pumping source can dominate the pumping dynamics \cite{Kalinikos1997,Gulyaev1997,pssr.202070022} and impact the selectivity of the spin-wave generation and amplification. Additionally, a decreased PP efficiency is expected in nano-sized structures with an aspect ratio close to one due to a vanishing ellepticity of the magnetization precession. Thus, an investigation of PP in nano-scaled conduits is of great interest to assess its potential future use for magnonic applications.

\begin{figure}[tbh]
    \includegraphics[width=84mm]{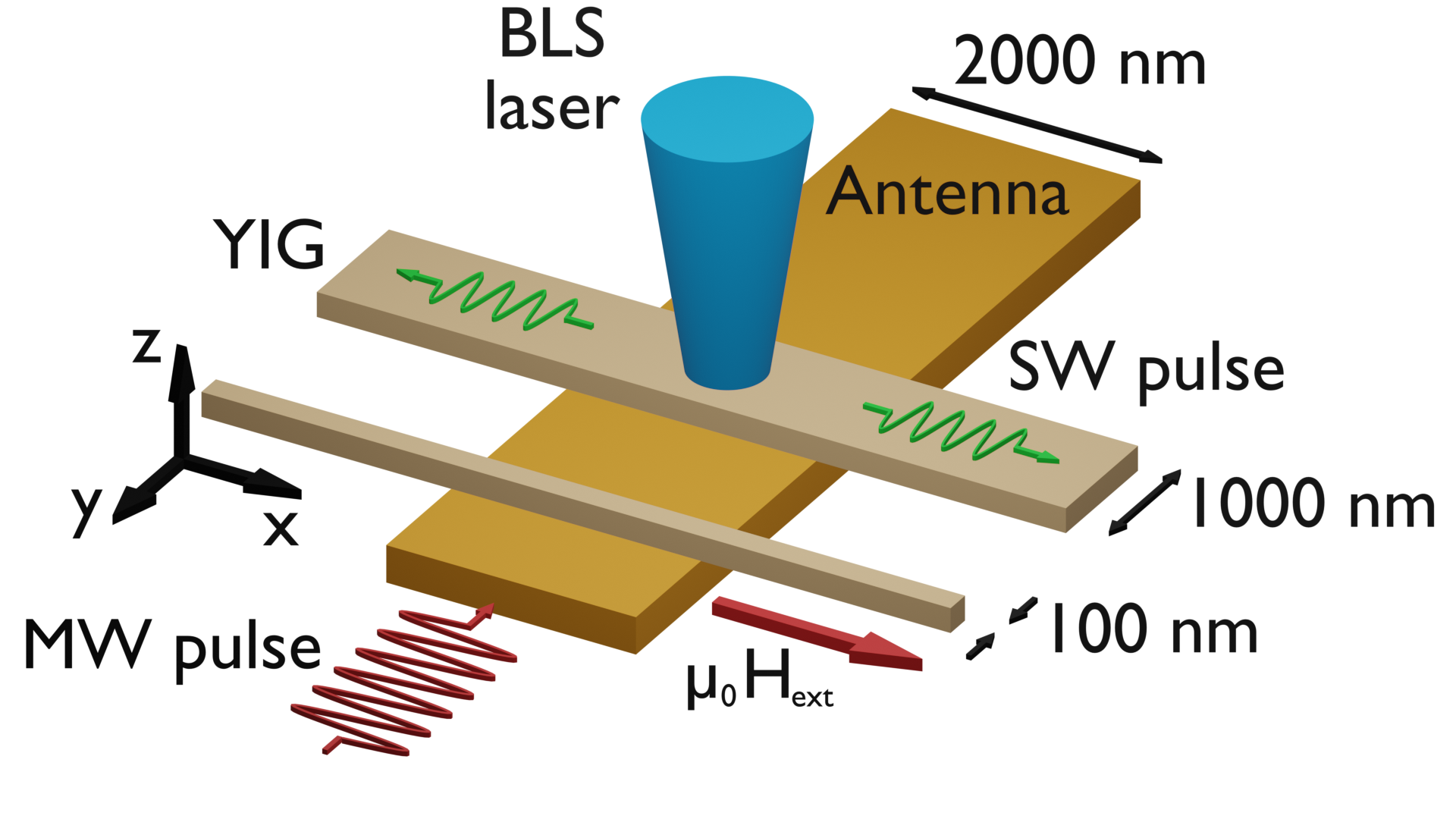}
    \caption{\label{Figure1} Schematic picture of the structure under investigation. A bias magnetic field $\mu_0 H_{\textrm{ext}}$ is applied longitudinal to the YIG conduits (\textit{x}-direction) and microwave-frequency pulses with a carrier frequency of \SI{7}{\giga\hertz} are fed into a perpendicularly oriented gold antenna to generate the pumping field. Micro-focused BLS scans are performed in the center of the pumping area and the respective conduit.}
\end{figure}
\section{Theoretical background}

Parallel pumping of spin waves is the interaction of a microwave pumping field $\mu_{0}h_{\text{p}}$ with a dynamic magnetization component, where both are oriented in parallel to the static magnetization of the system. Such a dynamic component arises due to an elliptic trajectory of the magnetization precession, caused by the presence of anisotropies (e.g., the shape anisotropy), in combination with the fact that the length of the magnetization vector is conserved \cite{BRACHER20171}. The pumping process can be described as the splitting of a microwave photon of the pumping field into two magnons, typically located at half the pumping frequency and with opposite wavevector (due to the negligible momentum of the photon and if the spatial confinement of the pumping field is weak). If the energy gain eventually exceeds the losses of the system, the spin-wave population will rise exponentially in time from the thermal spin-wave level. Thus, PP is a threshold process, while the point of compensation is referred to as the parametric instability threshold with the associated threshold field $\mu_{0}h_{\text{th}}$. A fundamental description of the PP dynamics is given in \mbox{Ref. \cite{Lvov1994}}, from which the parametric instability threshold is found as:
\begin{equation}\label{Eq1}
    \mu_{0}h^{\text{int}}_{\text{th}}=\frac{\Gamma_{\text{k}}}{V_{\text{k}}}.%
\end{equation}
Here, $V_{\text{k}}$ is the coupling parameter, which describes the efficiency of the parametric interaction and $\Gamma_{\text{k}}$ denotes the total intrinsic spin-wave relaxation rate of the pumped spin-wave group with wavevector $k$. The relaxation rate can be expressed as \cite{VerbaDamping2018}
\begin{equation}\label{Eq2}
    \Gamma_{\text{k}}=\left(\alpha\omega_{\text{k}}+\frac{\gamma\mu_{0}\Delta H_{0}}{2}\right)\epsilon_{\text{k}},%
\end{equation}
where $\alpha$ is the Gilbert damping parameter, $\omega_{\text{k}}=2\pi f_\textrm{k}$ the spin-wave frequency, $\gamma$ the gyromagnetic ratio, $\mu_{0}\Delta H_{0}$ the inhomogeneous linewidth broadening, and $\epsilon_{\text{k}}$ an ellipticity coefficient derived from the spin-wave dispersion. \mbox{Equation \ref{Eq1}} only provides a valid description of the parametric instability threshold in the case of a non-local pumping field ("global pumping" of the system). In contrast, if the pumping field is spatially confined, spin waves can propagate out of the pumping zone, which constitutes an additional loss channel. These losses are referred to as radiative damping, while the losses included in \mbox{Eq. \ref{Eq2}} are referred to as intrinsic damping. The associated parametric instability threshold, in the case of a quasi one-dimensional system (nano-conduit) and negligible intrinsic losses, follows as \cite{Melkov_2001}:
\begin{equation}\label{Eq3}
     \mu_{0}h^{\text{rad}}_{\text{th}}=\frac{v_{\text{g}}}{LV_{\text{k}}}\frac{\arccos{(\alpha_{\text{n}}})}{\sqrt{1-\alpha^{2}_{\text{n}}}}.%
\end{equation}
Here, $\alpha_{\text{n}}=\lvert \text{sinc}(kL) \rvert$ is the parameter of non-adiabacity, $L$ the length of the pumping area and $v_{\text{g}}$ the spin-wave group velocity. In the general case, the parametric instability threshold is not the mere sum of intrinsic and radiative thresholds but instead has to be determined from the following implicit equation \cite{Kalinikos1997,Verba_2017,pssr.202070022}: 
\begin{eqnarray}\label{Eq4}
    \frac{\sqrt{(V_{\text{k}} \mu_{0}h_{\text{th}})^2-(\Gamma_{\text{k}}-\alpha_{\text{n}}V_{\text{k}} \mu_{0}h_{\text{th}})^2}}{\Gamma_{\text{k}}-\alpha_{\text{n}}V_{\text{k}} \mu_{0}h_{\text{th}}}=\nonumber
\end{eqnarray}
\begin{eqnarray}
    -\tan{\left(\frac{\sqrt{(V_{\text{k}} \mu_{0}h_{\text{th}})^2-(\Gamma_{\text{k}}-\alpha_{\text{n}}V_{\text{k}} \mu_{0}h_{\text{th}})^2}L}{v_{\text{g}}}\right)}.
\end{eqnarray}
Under certain circumstances, a non-resonant ($\omega_{\text{k1}}=\omega_{\text{k2}}\neq\frac{\omega_{\text{p}}}{2}$, with pumping frequency $\omega_{\text{p}}=2\pi f_\textrm{p}$) \cite{Melkov_2013,Verba_2021} or even a non-degenerate non-resonant ($\omega_{\text{k1}}\neq\omega_{\text{k2}}\neq\frac{\omega_{\text{p}}}{2}$) splitting of the pumped spin waves can be favorable \cite{Melkov_2006,VerbaPump_2015}. In the former case, the parametric instability threshold of \mbox{Eq. \ref{Eq1}} is modified as follows \cite{BRACHER20171}:  
\begin{equation}\label{Eq5}
     \mu_{0}h^{\text{int}}_{\text{th}}=\sqrt{\frac{\Gamma_{\text{k}}^2+(\omega_{\text{k}}-\frac{\omega_{\text{p}}}{2})^2}{V_{\text{k}}^2}}.%
\end{equation}
Please note, a description of the dynamics beyond the instability threshold in the regime of small supercriticalities ($h_{\text{p}}/h_{\text{th}}$) is given in \mbox{Ref. \cite{Lvov1994}}, while for large supercriticalities, a strong nonlinear shift of the spin-wave dispersion and additional multi-magnon scattering effects have to be taken into account.

\section{Experimental setup}
\begin{figure*}[tbh]
    \includegraphics{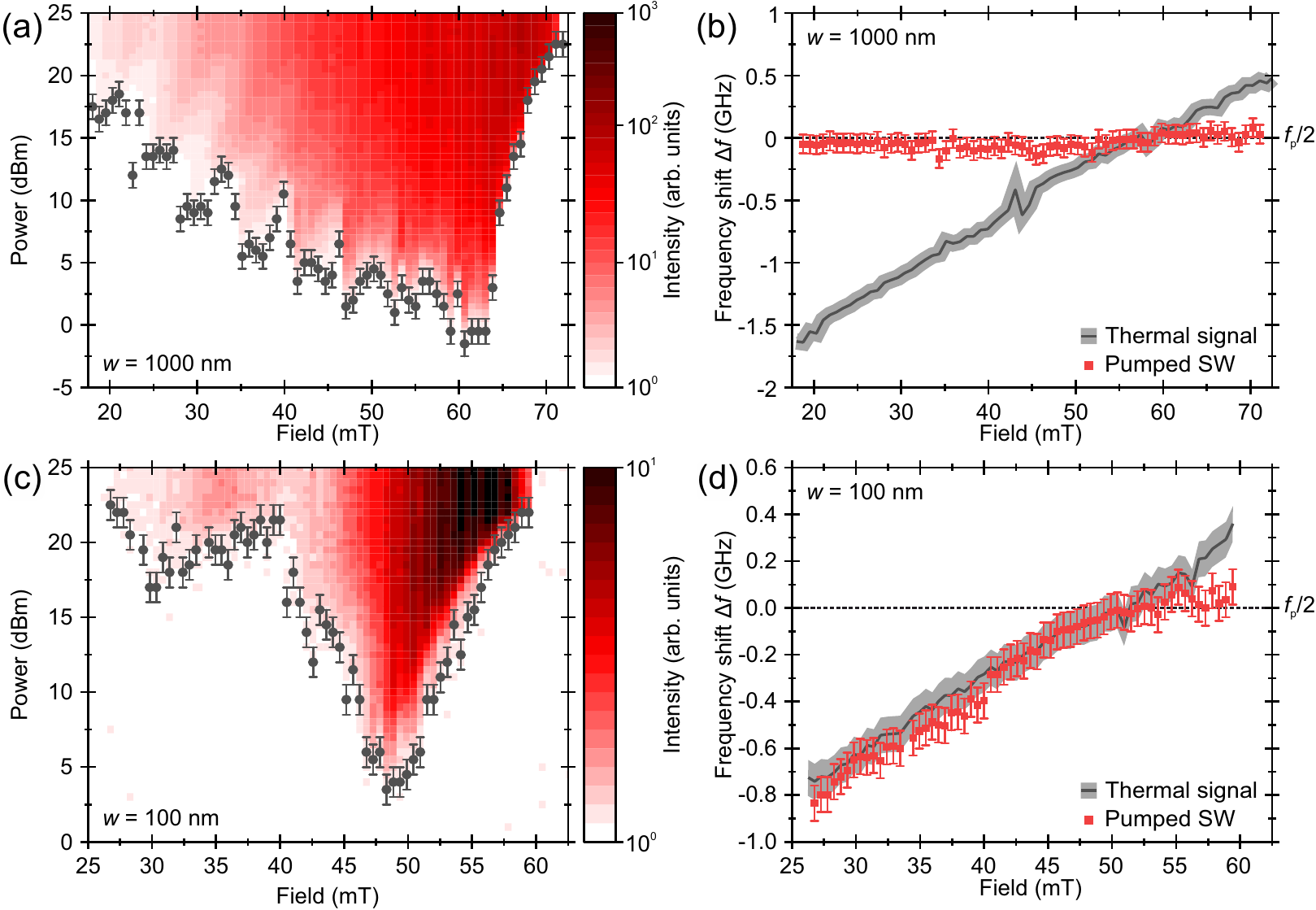}
    \caption{\label{Figure2} (a) and (c) Normalized pumping spectra for a respective conduit width of $w=\SI{1000}{\nano\metre}$ and $w=\SI{100}{\nano\metre}$. Intensity displayed in log-scale. Black dots: extracted parametric instability threshold. The spectra are normalized with respect to the thermal spin-wave level. (b) and (d) Extracted frequency of the generated spin waves (red dots) of (a) and (c) relative to $f_{\text{p}}/2$ for a small supercriticality. Solid black line: frequency of the thermally excited spin waves (error bar displayed as shaded area).}
\end{figure*}
In this study, a thin \SI{44}{\nano\metre}-thick lanthanum-doped \hkl(111) YIG film is used, which was grown on top of a \SI{500}{\micro\metre}-thick gadolinium gallium garnet substrate by liquid phase epitaxy \cite{Dubs_2020}. The plain film parameters were measured using vector network analyzer ferromagnetic resonance spectroscopy (VNA-FMR) \cite{MAKSYMOV2015253} and micro-focused Brillouin-light-scattering (BLS) spectroscopy \cite{ThomasS_2015}, revealing an saturation magnetization of $M_{\textrm{s}}=$\SI[separate-uncertainty = true]{140.7(28)}{\kilo\ampere\per\metre}, a Gilbert damping parameter of $\alpha =$\SI[separate-uncertainty = true]{1.75(8)e-4}, an inhomogeneous linewidth broadening of $\mu_0 \Delta H_0=$\SI[separate-uncertainty = true]{0.18(1)}{\milli\tesla} and an exchange constant of $A_{\textrm{ex}}=$\SI[separate-uncertainty = true]{4.22(21)}{\pico\joule\per\metre}. Subsequently, micro and nano-sized waveguides with a respective conduit width of $w=\SI{1000}{\nano\metre}$ and $w=\SI{100}{\nano\metre}$ were fabricated using a hard-mask ion beam milling procedure \cite{Heinz_2020}. A titanium/gold single-strip antenna was added on top perpendicular to the long axis of the waveguides, with a thickness of \SI{10}{\nano\metre}/\SI{150}{\nano\metre} and a width of $L=\SI{2}{\micro\metre}$ [see \mbox{Fig. \ref{Figure1}(a)}]. In the experiment, a bias magnetic field $\mu_0 H_{\textrm{ext}}$ is applied along the conduit's long axis, while microwave-frequency current pulses (pumping pulses) with a pulse length of \SI{50}{\nano\second} and a repetition time of \SI{280}{\nano\second} are fed into the antenna. The in-plane component of the resulting oscillating Oersted field of the antenna is oriented longitudinally to the conduits (in parallel to the internal static magnetization) and can thus act as the driving field for the PP process. The generated spin waves are investigated employing micro-focused BLS spectroscopy by focusing a single-frequency laser with a wavelength of \SI{457}{\nano\metre} and  \SI{2}{\milli\watt} effective laser power through the substrate of the sample on the center of the pumping area using a compensating microscope objective (magnification $100\times$, numerical aperture $\textrm{NA} = 0.85$). In-plane spin-wave wavevectors up to $k\approx\SI{24}{\radian\per\micro\metre}$ can be detected. To eliminate the influence of a varying microwave transmission characteristic of both the antenna and the used equipment, the frequency of the pumping pulse is fixed to $f_\textrm{p}=\SI{7}{\giga\hertz}$ and the external magnetic field is varied instead.\newline

\section{Results and discussions}

\subsection{Parallel pumping in micron-sized systems}

\mbox{Figure \ref{Figure2}(a)} shows the spin-wave intensity as a function of the external magnetic field and the applied pumping power, referred to as the pump spectrum in the following, for a conduit width of $w=\SI{1000}{\nano\metre}$. The extracted parametric instability threshold, displayed as the full black dots, resembles a commonly observed curve which is known as the "butterfly curve" and can be understood by taking the dependencies of the intrinsic and radiative losses into account, see \footnote{See Supplemental Material for the calculated parameters of the system (group velocity, relaxation time, ellipticity coefficient and coupling parameter) and the respective thresholds, for a micro-magnetic simulation of the spin-wave dispersion relationship, for the calculated Fourier spectrum of the pumping pulse, for micro-magnetic simulations of the pumping process, for the extracted order of the power law for the \SI{100}{\nano\metre}-wide conduit, and for an exemplary approximation of the inverse spin-wave intensity rise time and the calculated relaxation time assuming the full linewidth contribution or only Gilbert-type relaxation.}. The coupling parameter is basically a measure for the ellipticity of the magnetization precession, which decreases with increasing longitudinal wavevector $k_\text{x}$ and with increasing width-mode order $n$, thus, resulting in intrinsic threshold minima for the respective $k_\text{x}^n=0$ points and an increasing threshold for smaller magnetic fields. Similarly, the radiative losses are minimized when the group velocity vanishes, which is the case at the respective mode's band bottom. Thus, the dipolar regime of higher-order width modes features a smaller threshold than the exchange regime of lower order modes, which causes the distinct peculiarities and jumps observed in the butterfly curve. For magnetic fields $\mu_0 H_{\textrm{ext}}>\SI{63}{\milli\tesla}$, a cut-off of the spin-wave manifold occurs, since the band bottom of the fundamental mode lies above $f_{\text{p}}/2=\SI{3.5}{\giga\hertz}$, which renders a resonant pumping process not possible leading to a sharp increase of the threshold. Still, parallel pumping is observed which can be attributed to several effects. In this region, a hard type of parametric excitation can be realized, driven by a nonlinear frequency shift of the spin-wave dispersion relation \cite{Verba_2017}. A signature of this hard excitation is a large spin-wave amplitude at the threshold, which can be clearly seen in \mbox{Fig. \ref{Figure2}(a)}. Additionally, at large excitation powers, the dispersion relationship can be shifted by a reduction of the effective magnetization due to Joule heating of the pumping pulse, as well as by a forced excitation driven by the out-of-plane field component of the pumping field.\newline
To specify the observed dynamics, the relative frequency shift of the pumped spin waves $\Delta f=f_\textrm{k}-f_\textrm{p}/2$ is extracted and shown in \mbox{Fig. \ref{Figure2}(b)}. In addition, the frequency of the thermally excited spin waves of the fundamental mode ($n=1$) is shown, which corresponds to the dispersion's dipolar regime of small wavevectors. As expected, in the case of a small supercriticality, the parametric excitation always takes place at half the pumping frequency, even in the cut-off regime of the spin-wave manifold, thus, confirming the aforementioned considerations of a dispersion shift in this regime.\newline 
The presented findings are in agreement to recent studies (see Ref. \cite{pssr.202070022}), and, in the following, provide an important reference basis to assess and evaluate the dynamics of parallel pumping in the \SI{100}{\nano\metre}-wide nano-conduit.

\subsection{Parallel pumping in magnonic nano-conduits}
\begin{figure}
    \includegraphics{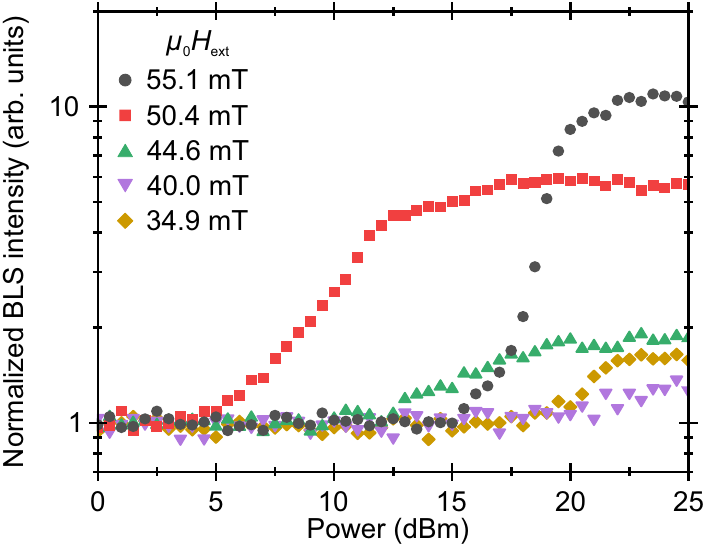}
    \caption{\label{Figure3} Normalized threshold curves of \mbox{Fig. \ref{Figure2}(c)} for selected magnetic fields. Normalization with respect to the thermal spin-wave level.}
\end{figure}
In \mbox{Fig. \ref{Figure2}(c)} the pump spectrum is shown for the \SI{100}{\nano\metre}-wide nano-conduit. In comparison to \mbox{Fig. \ref{Figure2}(a)} the shape of the butterfly curve is noticeably different, as it does not exhibit several minima, but only one distinct threshold minimum. This is in agreement with the expectation, since the nano-conduit effectively represents a single-mode system as any higher order width or thickness modes are significantly lifted in frequency due to the strong quantization of the system, see \cite{Note1}. The behaviour in the cut-off regime for magnetic fields $\mu_0 H_{\textrm{ext}}>\SI{50.5}{\milli\tesla}$ is very similar to \mbox{Fig. \ref{Figure2}(a)}, while the threshold increases continuously in the regime from $\mu_0 H_{\textrm{ext}}=\SI{45}{\milli\tesla} - \SI{40}{\milli\tesla}$, until the spin-wave intensity nearly vanishes for the maximal applied power. For even smaller magnetic fields a reduced threshold is observed and a noticeable spin-wave intensity is detected again.\newline
The pumped spin waves above the cut-off of the spin-wave manifold are located at $f_{\text{p}}/2$, but for smaller magnetic fields, the spin waves are significantly shifted to smaller frequencies, contrary to the case of the \SI{1000}{\nano\metre}-wide waveguide [see \mbox{Fig. \ref{Figure2}(d)}]. Here, a non-resonant pumping process is observed, which effectively pumps spin waves into the dipolar regime of the dispersion. In principle, such a frequency detuning is possible within the frequency uncertainty of the pumping pulse. However, this effect can be estimated to be on the order of $\pm\SI{50}{\mega\hertz}$, see \cite{Note1}, which is insufficient to cause the experimental observation. In general, a non-resonant process is unfavorable in comparison to the resonant generation of spin waves since it corresponds to an increase of the effective intrinsic threshold, as it can be seen from \mbox{Eq. \ref{Eq5}}. Still, it can potentially be driven by a minimization of the radiative losses, which nearly vanish in the dipolar regime of the dispersion, see \cite{Note1}. Since the investigated case of non-resonant non-adiabatic pumping in nano-sized structures is not covered within current theoretical models, we perform mico-magnetic simulations of the pumping process using the MuMax3 framework \cite{MuMAX3}. In contrast to the experimental results, the simulations show only a resonant pumping of exchange spin waves (see \cite{Note1}). However, one needs to consider that micro-magnetic simulations only account for viscous Gilbert-type damping and radiative losses, while being incapable to account for non-Gilbert type relaxation mechanisms, which can contribute significantly to the system's relaxation in high-quality YIG films \cite{LeCraw_1961}. Thus, the simulations cannot necessarily describe non-resonant pumping processes as observed in the experiment.\newline
Furthermore, to exclude a linear excitation process as the underlying mechanism of the experimental observation, selected threshold curves of \mbox{Fig. \ref{Figure2}(c)} are exemplary shown in \mbox{Fig. \ref{Figure3}}. A clear threshold behavior is observed, while extraction of the order $m$ of the associated power law ($Intensity\propto Power^m$) reveals $m\neq1$, see \cite{Note1}. Thus, any type of direct linear excitation can be excluded.\newline

\subsection{Limitations of parallel pumping in magnonic conduits}
In the following, we discuss the limiting factors of PP regarding the generation of spin waves in magnonic conduits. While in micron-sized conduits a variety of different (even and odd) width modes can be easily selected and amplified, the resonant generation of exchange spin waves is restricted by the presence of dipolar spin-wave states of other width modes, which will always feature a smaller effective threshold (see \cite{Note1}). In this regard, the effective single-mode property of nano-sized conduits is very beneficial, but non-resonant pumping processes as observed above seem to limit the versatility of PP. Thus, the development of advanced theoretical models is required to unravel the underlying mechanisms and to evaluate how the tuning of system parameters, e.g., the size of the pumping area, can be used to effectively model the accessible wavevector regime. We would like to point out that a pre-existing coherent spin-wave population, e.g., externally provided by a 2\textsuperscript{nd} antenna, significantly impacts the dynamics of PP, since the effective energy gain is proportional to the pumped spin-wave group's population \cite{BracherPP_2014,BRACHER20171}. In this regard, the resonant amplification of exchange spin waves in nano-sized conduits might be favorable over the non-resonant generation of dipolar spin waves.\newline
An additional limitation arises for nano-sized conduits approaching an aspect ratio of one, which is the significant decrease of the coupling strength due to the unpinning of the spin-wave modes \cite{WangPinning_2019}. With respect to the \SI{1000}{\nano\metre}-wide conduit and for $n=1$ and $k_\text{x}=0$, the ellipticity coefficient drops to \mbox{$\epsilon_\mathrm{k}(w=\SI{100}{\nano\meter})/\epsilon_\mathrm{k}(w=\SI{1000}{\nano\meter})=\SI{86.1}{\percent}$} and the coupling parameter drops significantly to \mbox{$V_\mathrm{k}(w=\SI{100}{\nano\meter})/V_\mathrm{k}(w=\SI{1000}{\nano\meter})=\SI{31.4}{\percent}$}. In this regard, a reasonably small threshold is found for the \SI{100}{\nano\metre}-wide conduit rendering PP feasible in such systems.\newline

\subsection{Effective spin-wave relaxation time}
In the following, we utilize PP not just as a mere tool to generate spin waves, but to gain access to the effective spin-wave relaxation time $\tau_{\text{eff}}$ (similar to Ref. \cite{PhysRevB.84.094401}) of a nano-sized conduit. In fact, PP provides direct access to this parameter rendering it superior in comparison to other methods, e.g., based on group velocity and decay length measurements. The time evolution of the spin-wave intensity $I$ for small supercriticalities can be described using a simple exponential function \cite{BRACHER20171}:
\begin{equation}\label{EqIntenTime}
    I(t)=I_0 \exp{\left(\frac{t}{\tau_{\text{rise}}}\right)}.%
\end{equation}
Here, $I_0$ denotes the inital spin-wave intensity (thermal population) and $\tau_{\text{rise}}$ is the effective rise time, which, in the case of dominating intrinsic losses, follows as \cite{BRACHER20171}:
\begin{equation}\label{EqRise}
    \tau_{\text{rise}}^{-1}=2\mu_0 h_{\textrm{p}}V_{\text{k}}-2\Gamma_{\text{eff}}=\sqrt{P}V_{\text{k}}^{'}-2\tau_{\text{eff}}^{-1}.%
\end{equation}
\begin{figure}
    \includegraphics{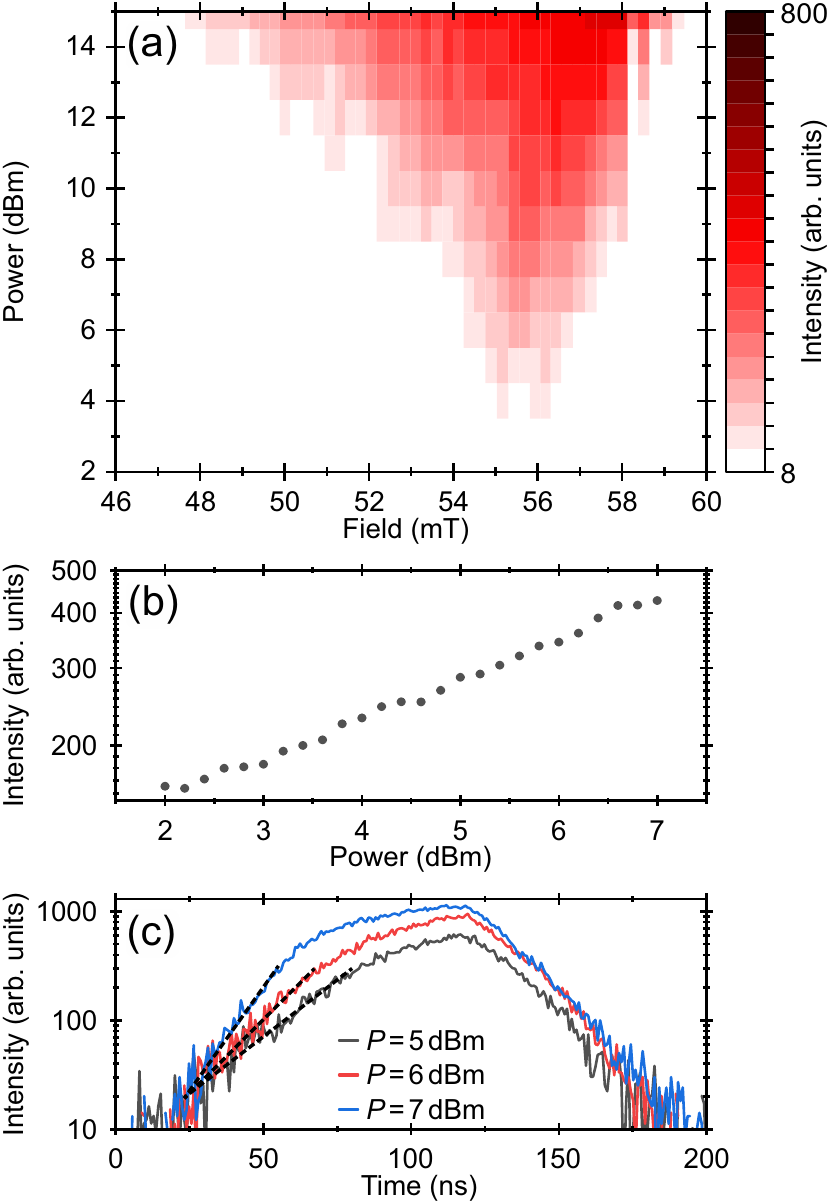}
    \caption{\label{Figure4} (a) Pump spectrum for a conduit width of $w=\SI{100}{\nano\metre}$. Pumping pulse parameters are changed to \SI{100}{\nano\second} pulse length and \SI{350}{\nano\second} repetition time and laser power increased to \SI{5}{\milli\watt}, causing a shift of the dispersion compared to \mbox{Fig. \ref{Figure2}(c)}. Intensity displayed in log-scale. (b) Threshold curve and (c) time traces of the generated spin-wave pulses for different applied microwave powers $P$ and $\mu_0 H_{\textrm{ext}}=\SI{55.4}{\milli\tesla}$. Dashed lines: approximation of the rising edges.}
\end{figure}
\begin{figure}
    \includegraphics{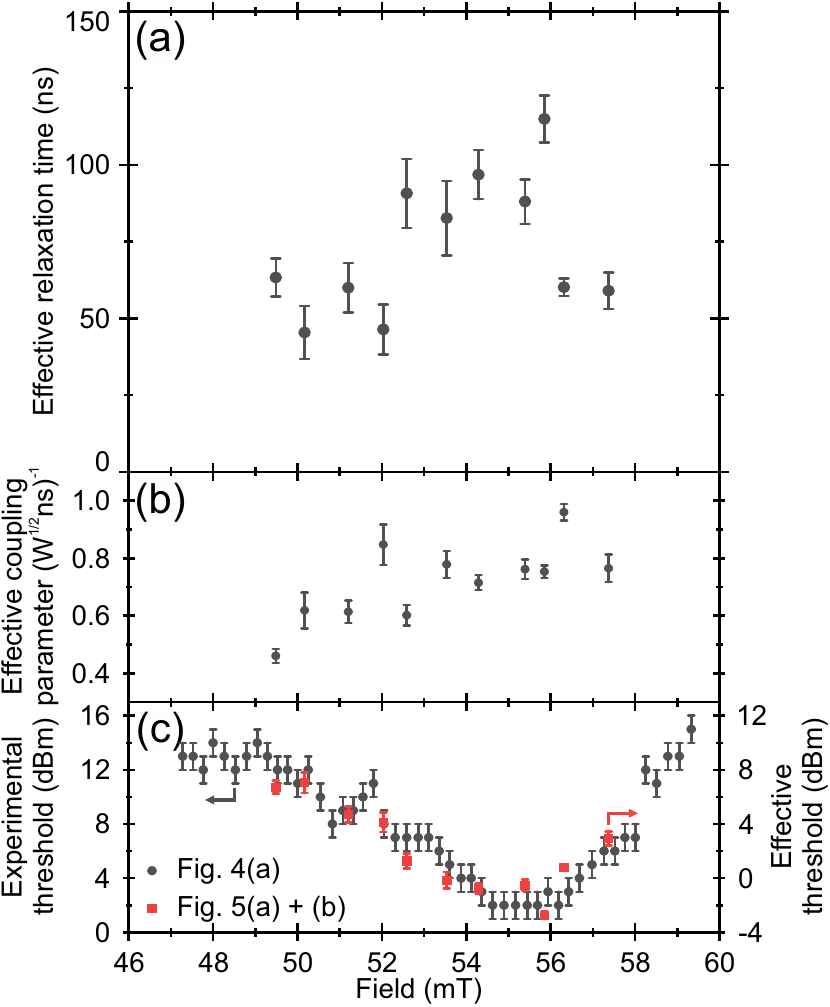}
    \caption{\label{Figure5} (a) Effective spin-wave relaxation time and (b) effective coupling parameter as a function of the external magnetic field. (c) Effective threshold calculated from (a) and (b) in comparison to the experimental threshold of \mbox{Fig. \ref{Figure4}(a)}.}
\end{figure}
The pumping field $\mu_0 h_{\textrm{p}}$ can be considered as proportional to the square root of the applied pumping power $P$, while any proportionality constant is absorbed into the effective coupling parameter $V_{\text{k}}^{'}$. This allows to extract the effective relaxation time from a power dependent set of time traces, see \cite{Note1}. Please note, these considerations are only valid if the radiative losses can be neglected or be considered much smaller than the intrinsic losses, which, for nano-sized conduits, holds true in the regime of small wavevectors. In the following, the applied pumping pulses are changed to \SI{100}{\nano\second} pulse length and \SI{350}{\nano\second} repetition time to provide a sufficient observation time of the rising spin-wave intensity. In addition, the laser power is increased to \SI{5}{\milli\watt}, which causes a shift of the dispersion compared to \mbox{Fig. \ref{Figure2}(c)} due to increased microwave and laser heating. \mbox{Figure \ref{Figure4}(a)} shows the pump spectrum of the \SI{100}{\nano\metre}-wide conduit and \mbox{Fig. \ref{Figure4}(b)} exemplary a threshold curve of the spin-wave intensity for $\mu_0 H_{\textrm{ext}}=\SI{55.4}{\milli\tesla}$. Within the investigated power range, the supercriticalitiy can be regarded as small since the threshold curve does not saturate considerably. However, a change of slope is observed in the rising edge of the spin-wave time traces [see \mbox{Fig. \ref{Figure4}(c)}], thus, the approximation of the rising edges according to \mbox{Eq. \ref{EqIntenTime}} is only considered up to this intensity level. In \mbox{Fig. \ref{Figure5}(a) and (b)}, the effective relaxation time and the effective coupling parameter extracted according to \mbox{Eqn. \ref{EqIntenTime} and \ref{EqRise}} are shown. We find a large relaxation time up to \SI[separate-uncertainty = true]{115.0(76)}{\nano\second} in the regime of the dispersion band bottom. This is significantly larger than the theoretical prediction assuming a full contribution of the inhomogeneous linewidth broadening as a loss channel, but smaller than a pure Gilbert-type relaxation, see \cite{Note1}. Please note, the plain film parameters extracted by VNA-FMR are likely to overestimate the effective local inhomogeneity \cite{Hahn_2014} and a potential influence of the structuring procedure cannot be excluded. Nonetheless, the extracted relaxation time is much larger than values found in microscopic or macroscopic systems of other commonly used magnonic materials such as Permalloy \cite{ShichiPermalloy2015}. The results for smaller magnetic fields have to be considered with caution, since the pumping is expected to be slightly non-resonant as found above. In fact, the spin waves are always located in the dipolar regime of the dispersion, thus, the radiative losses can be considered as small, but the additional frequency shift still influences the evaluation. A trust check can be made by recalculating the effective threshold from the relaxation time and the coupling parameter, which is shown in \mbox{Fig. \ref{Figure5}(c)}. The threshold fits well to the experimental threshold extracted from \mbox{Fig. \ref{Figure4}(a)}, if accounted for the unknown offset in power due to impedance mismatch of the antenna and losses in the equipment. Thus, the effective spin-wave relaxation time found above can be regarded as valid.\newline

\subsection{Dynamics in the regime of large supercriticality}\label{sec:super}
\begin{figure}
    \includegraphics{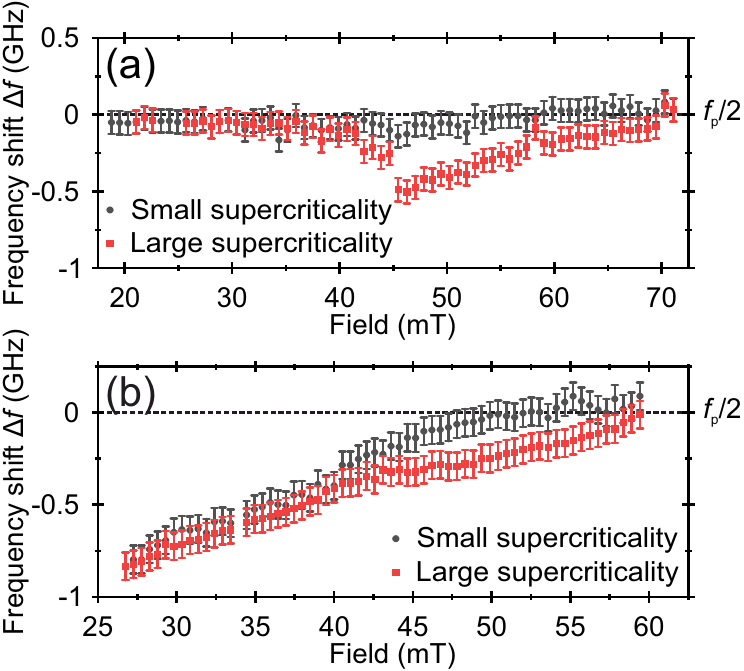}
    \caption{\label{Figure6} Relative frequency shift of the pumped spin-wave group in comparison for small and large supercriticalities, for (a) $w=\SI{1000}{\nano\metre}$ and (b) $w=\SI{100}{\nano\metre}$.}
\end{figure}
\begin{figure*}[htb!]
    \includegraphics{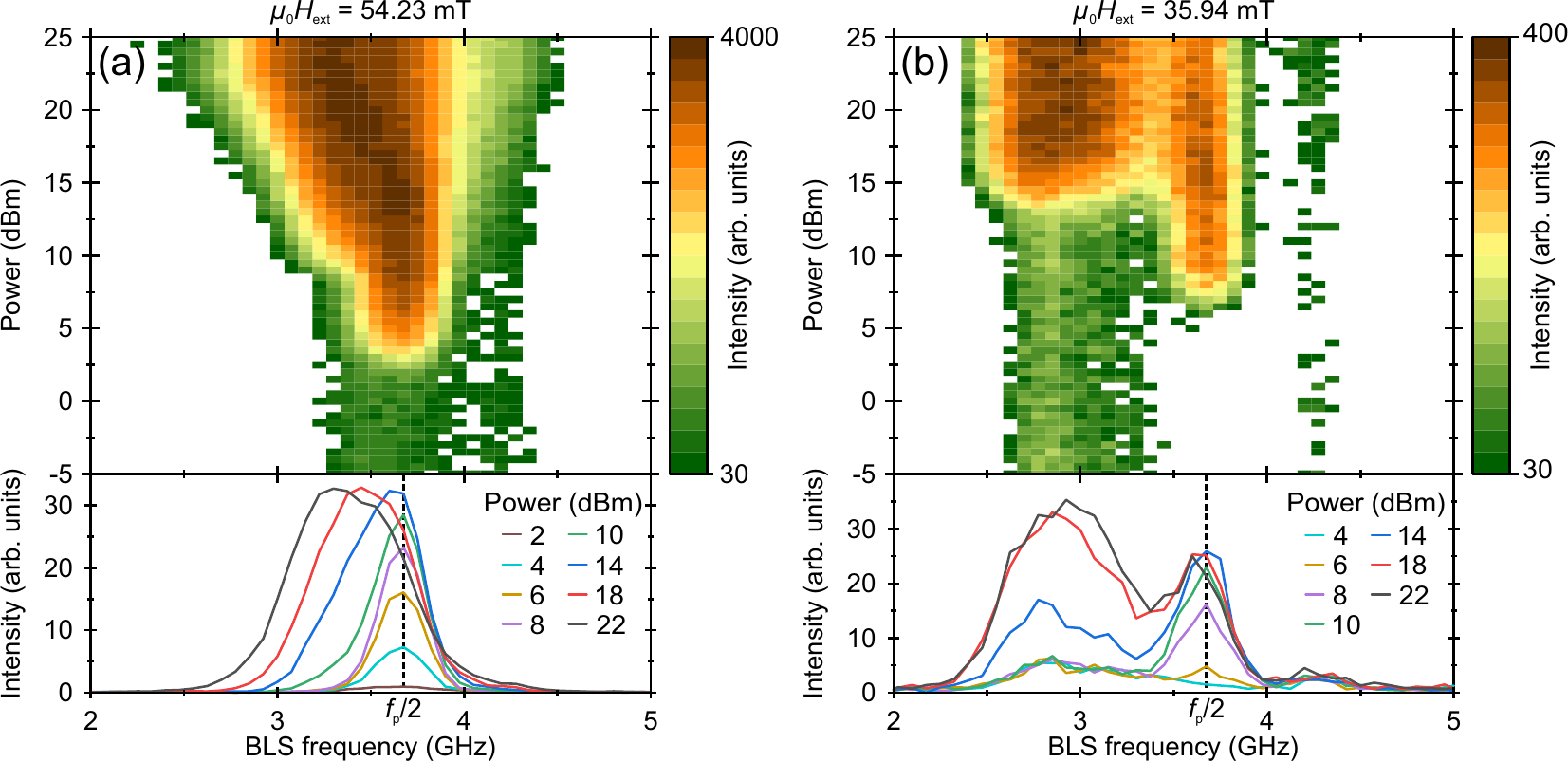}
    \caption{\label{Figure7} Power dependent BLS-spectra for (a) $\mu_0 H_{\textrm{ext}}=\SI{54.23}{\milli\tesla}$ and (b) $\mu_0 H_{\textrm{ext}}=\SI{35.94}{\milli\tesla}$ for the $w=\SI{1000}{\nano\metre}$-wide conduit. Please note, the method of BLS-spectroscopy induces a small frequency shift, thus, $f_{\text{p}}/2=\SI{3.675}{\giga\hertz}\neq\SI{3.5}{\giga\hertz}$ in the displayed spectra (marked by the dashed black line).}
\end{figure*}
In the following, the regime of large supercriticalities and the additional processes which are observed to take place are discussed. In \mbox{Fig. \ref{Figure6}(a)} the relative frequency shift of the pumped spin waves is shown for the \SI{1000}{\nano\metre}-wide conduit. We observe a deviation from $f_{\text{p}}/2$ for a large supercriticality and $\mu_0 H_{\textrm{ext}}>\SI{42}{\milli\tesla}$. In fact, a non-resonant pumping process to frequencies $f_{\text{k}}<f_{\text{p}}/2$ occurs, caused by the large spin-wave population achieved. This becomes evident from the associated spin-wave spectra, exemplary shown for $\mu_0 H_{\textrm{ext}}=\SI{54.23}{\milli\tesla}$ in \mbox{Fig. \ref{Figure7}(a)}. For large magnetic fields, the spin-wave intensity will rise for powers above the threshold, but the spin waves are located at $f_{\text{p}}/2$. At very large powers above the threshold, the pumped spin-waves shift in frequency, driven by the associated increasing nonlinear frequency shift. In addition, a further increase of the intensity is limited, likely by an induced dephasing of the pumped spin waves and the pumping field, which reduces the effective energy gain of the system \cite{BRACHER20171}. In contrast, for external fields $\mu_0 H_{\textrm{ext}}<\SI{42}{\milli\tesla}$, the frequency of the pumped spin waves relaxes to $f_{\text{p}}/2$ [see \mbox{Fig. \ref{Figure6}(a)}] and instead of a non-resonant pumping process, nonlinear scattering processes populating the dipolar regime of the dispersion are observed. This is exemplary shown in \mbox{Fig. \ref{Figure7}(b)} for $\mu_0 H_{\textrm{ext}}=\SI{35.94}{\milli\tesla}$, where the pumped spin waves are always found at $f_{\text{p}}/2$, even for the maximal supercriticality. However, the spin-wave population at $f_{\text{p}}/2$ saturates at a certain power, at which a sudden increase of the spin-wave population at lower frequencies is detected. This is a strong indication for the onset of a multi-magnon scattering process, which causes a population of the dipolar regime. However, due to the conservation of energy, three-magnon-splitting processes are excluded and necessarily spin waves have to be scattered to larger frequencies. These spin waves are not observed in the experiment since they are likely situated beyond the wavevector detection limit of the used setup.\newline
Please note, \mbox{Fig. \ref{Figure6}(b)} also shows the case of large supercriticalities for the \SI{100}{\nano\metre}-wide conduit. Here, the pumping is already non-resonant for small supercriticalities and larger applied powers are only observed to introduce an additional frequency shift in the regime of $\mu_0 H_{\textrm{ext}}>\SI{40}{\milli\tesla}$.  Multi-magnon scattering cannot take place since the spin waves already populate the lowest energy states of the system and, thus, energy conservation cannot be fulfilled for any scattering processes into higher frequencies. Only a three-magnon confluence process would be possible for a specific external field but, although the pump spectrum saturates considerably for large supercriticalities, we find no trace of such a process within the measured spectra.\newline

\section{Conclusion}
To conclude, we presented an investigation of parallel pumping processes in yttrium iron garnet magnonic conduits. In micron-sized structures, a variety of different (even and odd) width modes can be amplified but the amplification of exchange spin waves is restricted due to the presence of the dipolar regime of higher order width modes in the pumped frequency range. In contrast, nano-scaled conduits effectively provide a single-mode system, rendering them superior to micron-sized systems in this regard. However, we observe non-resonant pumping processes to be favorable over the resonant generation of exchange spin waves, leading to a population of the dipolar regime in an extended regime of the pump spectrum, which is likely driven by a minimization of the radiative losses. Our results emphasize that the development of an advanced theoretical model including viscous and non-viscous damping channels alike is required to describe the investigated regime of non-resonant non-adiabatic parallel pumping adequately. Furthermore, despite the significant decrease of the coupling strength in the investigated nano-conduit (drop of the coupling parameter to \mbox{$V_\mathrm{k}(w=\SI{100}{\nano\meter})/V_\mathrm{k}(w=\SI{1000}{\nano\meter})=\SI{31.4}{\percent}$} for $n=1$ and $k_\text{x}=0$), a reasonably small threshold was found rendering parallel pumping feasible even in nano-scaled systems. Thus, this work provides valuable guidelines for the future utilization of parallel pumping for an efficient spin-wave amplification on the nanoscale, which is key to the development of advanced magnonic networks.\newline

\begin{acknowledgments}
This research has been funded by the European Research Council project ERC Starting Grant 678309 MagnonCircuits, by the Deutsche Forschungsgemeinschaft (DFG, German Research Foundation) - 271741898, by the Collaborative Research Center SFB/TRR 173-268565370 (Project B01), and by the Austrian Science Fund (FWF) through the project I 4696-N. B.H. acknowledges support from the Graduate School Material Science in Mainz (MAINZ). R. V. acknowledges support by the National Research Foundation of Ukraine (Grant No. 2020.02/0261) and by the Ministry of Education and Science of Ukraine (project 0121U110090). The authors thank Burkard Hillebrands for support.
\end{acknowledgments}

\bibliography{PumpingBib}

\end{document}